\documentclass[twocolumn,pre,superscriptaddress,bibnotes,notitlepage,nofootinbib]{revtex4-1}
\usepackage{color}
\definecolor{red}{rgb}{0.75,0,0}
\definecolor{blue}{rgb}{0,0,0.75}
\definecolor{green}{rgb}{0,0.5,0}
\usepackage{ulem}
\usepackage{amsmath}
\usepackage{amssymb}
\usepackage{bm}
\usepackage{mathtools}
\usepackage{graphicx}
\usepackage{hyperref}
\usepackage{verbatim}
\usepackage{graphicx}
\usepackage{epsf,epstopdf,wrapfig}
\usepackage{graphics}
\usepackage{wrapfig}
\usepackage{dcolumn}
\usepackage{amssymb}
\usepackage{bm}
\usepackage{epsfig}
\usepackage{psfrag}
\usepackage{color}
\usepackage[utf8x]{inputenc}
\usepackage{ucs}
\usepackage{amsmath}
\usepackage{amsfonts}
\usepackage{amssymb}
\usepackage{graphicx}
\usepackage{dcolumn}
\usepackage{bm}
\usepackage{latexsym}
\usepackage{hyperref}
\DeclareMathAlphabet{\mathcalligra}{T1}{calligra}{m}{n}

\def\be{\begin{equation}}
\def\ee{\end{equation}}
\def\bea{\begin{eqnarray}}
\def\eea{\end{eqnarray}}

\def\besub{\begin{subequations}}
\def\eesub{\end{subequations}}

\def\3dots{\:\raisebox{-0.5ex}{$\stackrel{\textstyle.}{:}$}\:}
\def\beq{\begin{equation}}
\def\eeq{\end{equation}}
\def\bea{\begin{eqnarray}}
\def\eea{\end{eqnarray}}

\def\bwd{\begin{widetext}}
\def\ewd{\end{widetext}}

\newcommand{\bsf}[1]{\textsf{\textbf{#1}}}

\newcommand{\SR}[1]{\textcolor{black}{#1}}

\newcommand{\AMN}[1]{\textcolor{black}{#1}}
\newcommand{\AMR}[1]{\textcolor{black}{#1}}

\begin{document}
\title{Nonmutual torques and the unimportance of motility for long-range order in two-dimensional flocks}
\author{Lokrshi Prawar Dadhichi}
\affiliation{Tata Institute of Fundamental Research, Centre for Interdisciplinary Sciences, Hyderabad 500 107, India}
\author{Jitendra Kethapelli}
\affiliation{International Centre for Theoretical Sciences, Tata Institute of Fundamental Research, Bengaluru 560 089, India}
\author{Rahul Chajwa}
\affiliation{International Centre for Theoretical Sciences, Tata Institute of Fundamental Research, Bengaluru 560 089, India}
\author{Sriram Ramaswamy}\thanks{Adjunct Professor, Tata Institute of Fundamental Research, Hyderabad}
\email{sriram@iisc.ac.in}
\affiliation{Centre for Condensed Matter Theory, Department of Physics, Indian Institute of Science, Bangalore 560 012, India}
\author{Ananyo Maitra}
\email{nyomaitra07@gmail.com}
\affiliation{Sorbonne Universit\'{e} and CNRS, Laboratoire Jean Perrin, F-75005, Paris, France}
\begin{abstract}
As the constituent \SR{particles} of a flock are polar \SR{and in a driven state, their interactions must in general be fore-aft asymmetric and non-reciprocal}. Within a model that explicitly retains the \SR{classical} spin angular momentum field of the \SR{particles we show that the resulting asymmetric contribution to interparticle torques, if large enough}, leads to a buckling instability of the flock. Precisely this asymmetry also yields a natural mechanism for a difference between the speed of advection of information along the flock and the speed of the flock itself, concretely establishing that the absence of detailed balance, and not merely the breaking of Galilean invariance, is crucial for this distinction. \AMR{To highlight this we construct a model of asymmetrically interacting spins fixed to lattice points and demonstrate that the speed of advection of polarisation remains non-zero.} We delineate the conditions on parameters and wavenumber for the existence of the buckling instability. Our theory should be consequential for interpreting the behaviour of real animal \SR{groups} as well as 
experimental studies of artificial flocks composed of polar motile rods on substrates.
\end{abstract}
\maketitle
\normalem

\section{Introduction}
In the classic models of flocking \cite{vicsek,tonertu} and much of the later literature \cite{otherflock} each agent is assumed to adjust its direction of motion to the mean of its neighbours including itself, plus a random error. It has recently become clear \cite{ISM, Loewen} that, on time- and length-scales relevant to observations on real bird flocks, the \AMR{inertial} dynamics of this reorientation must be explicitly taken into account, via \SR{a classical} spin angular momentum on which the aligning interaction acts as a torque \cite{comm1}. This inertial effect was shown \cite{ISM,silent,XY} to give rise, on intermediate length scales \cite{Ram_maz}, to turning waves reminiscent of those predicted for inertial flocks in fluids \cite{simhaSR} or rotor 
lattices \cite{chailub}. On the longest scales, where damping by the ambient medium overcomes inertia, the dynamics is effectively described by the Toner-Tu \cite{tonertu} equations. 

\begin{figure}
  \includegraphics[width=3 truecm]{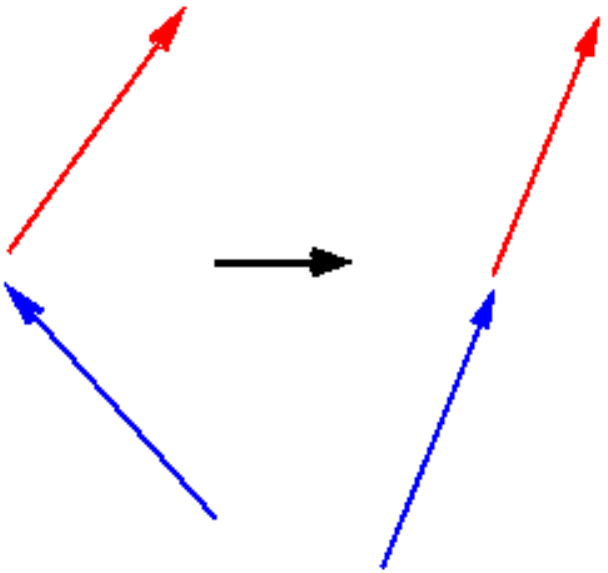}
  \caption{A pair of fore-aft-separated birds with misaligned velocities; for reasons of vision or airflow, the bird in front is slower to adjust its orientation. This \SR{asymmetric coupling} leads to a net rotation of the overall alignment of the birds.}
  \label{fig:chiral_rotate}
\end{figure}

However, in \cite{ISM,silent} the aligning field was implicitly taken to arise from an effective Hamiltonian, so the interactions were perfectly mutual. A pair of birds exerted opposing torques of equal magnitude on each other, conserving spin angular momentum. This is in principle unduly restrictive: interactions not governed by an energy function are permitted in systems out of thermal equilibrium \cite{suro, suro2,broken3rd},
and the dynamics takes place in contact with an ambient medium with which the birds can exchange both angular and linear momentum. Indeed, self-propulsion consists precisely in drawing linear momentum from the ambient medium, with directional bias determined by the structural polarity of the bird. \SR{In the flocking models we consider, birds are individually achiral and carry only a position and a vectorial orientation. Thus, despite the possibility of drawing \textit{angular} momentum from the ambient air, self-propelling activity will not lead to a net persistent rotational motion of an individual bird \cite{Ano_chi}. Transient chirality and, hence, spontaneous rotation can arise only through interaction between birds.} 
Consider a pair of birds flying one ahead of the other. The basic assumption of flocking models is that if the velocity vectors of the birds depart slightly from being parallel, an aligning torque arises. Each bird tries to rotate its motility direction to match that of its neighbour, but one expects in general that the aligning response of the leading bird to the trailing bird should be different from that of the trailer to the leader (Fig. \ref{fig:chiral_rotate}). Such \SR{asymmetry} of information transfer or response could arise, inter alia, from vision \cite{birdvisionwiki} or airflow \cite{jeb2011}. Our focus is distinct from that of ref. \cite{trigger} in which pairwise \SR{asymmetry}, distributed statistically across a collection of birds, gives rise to an ultra-rapid response and relaxation, and from \cite{Chate_asym} in which a spin-overdamped description is used from the start \cite{comm_bib}. In all three works, however, motility and \SR{asymmetric} aligning torques enter as two distinct manifestations of the nonequilibrium nature of the system. \AMR{Furthermore, our general mechanism for non-mutual interaction can arise even in \emph{non-motile} but active polar systems such as spins on a lattice.}

Here we consider this physically natural \SR{asymmetry} of interaction, within the continuum description of \cite{silent} flocks with turning inertia \cite{ISM}.
The interaction between birds flying precisely side-by-side is taken to be symmetric. \AMR{The \SR{asymmetric} interaction is an explicit microscopic nonequilibrium ingredient distinct from motility in this model. To highlight this, we construct a model of spins on a lattice which interact via non-mutual interactions and demonstrate that the equation of motion for the magnetisation is essentially equivalent to that of the velocity of motile flocks.
}
\AMR{We also consider the opposite situation -- a model in which} \SR{asymmetric} interaction is \AMR{\emph{not}} an explicit microscopic nonequilibrium ingredient -- \AMR{to demonstrate that asymmetric interactions may arise at the macroscopic level \emph{purely} due to a microscopic motility}. \AMR{We construct the dynamical equations of a system of} polar motile rods on a substrate, \AMR{\emph{explicitly taking the angular momentum density into account} \cite{Stark}}, and show that an \SR{antisymmetric contribution to the interaction matrix determining the reorienting torques emerges} purely due to motility and the tendency of polarisation in such systems to align with both an imposed mean flow and its gradient. \AMR{Taken together} this establishes that \SR{an asymmetric aligning interaction}, either emergent or due to an explicit microscopic interaction, is a generic \AMR{property of all active polar systems} \cite{tonertu, RMP, Chate_asym} and must be accounted for. \SR{Doing so, 
we find that} if the antisymmetric part $\mathcal{A}$ of the aligning interaction exceeds a threshold $|\mathcal{A}_c| \propto \sqrt{J}$ where $J$ is the symmetric part, \AMR{the} uniform \AMR{ordered states} undergo a spontaneous long-wavelength buckling instability with wavevector
along \SR{the mean direction of alignment of the flock.
}In other words, the \SR{effective longitudinal diffusivity for the relaxation of the orientation} of \AMR{the polar phase} generically turns negative signifying a destruction of the ordered \AMR{state}. \AMN{We demonstrate this by constructing the hydrodynamic equation for the velocity \AMR{(or polarisation)} field alone, after systematically eliminating the fast spin angular momentum field, obtaining an equation that has the same form as the Toner-Tu model, as required by symmetry, but with a longitudinal diffusivity that changes sign at a critical $|\mathcal{A}_c|$.} \SR{While one may directly examine the effect of a negative} longitudinal diffusivity in the Toner-Tu model, which is the spin overdamped limit of our model, our examination here, retaining the spin angular momentum of the flock explicitly, provides a physically appealing mechanism via which this may happen and connects it to an effective or explicit antisymmetric \SR{part of the} interaction between the \AMR{microscopic polar units}. \AMN{We calculate the full dispersion relation implied by our model when there is a small wavevector instability, and demonstrate that while the imaginary part of the eigenfrequency does turn negative at large enough wavevectors or small enough scales, this does not lead to a mechanism for wavevector selection unlike other active models with a diffusive instability \cite{Ano_apol, Aran_PRX, Dunkel} -- the dispersion relation features an extended \emph{wavevector-independent} plateau before decaying at larger wavevectors. }

Before showing how we arrived at these results, a remark is in order regarding advective effects in flocking models. In the Toner-Tu \cite{tonertu} equation the velocity or the polarisation field is not advected at the same rate as the density. While it is tempting to attribute this asymmetry simply to the absence of Galilean invariance in the theory, both Galilean invariance {\it and detailed balance} have to be absent for this effect to emerge, as was demonstrated in \cite{LPDJSTAT}. For a stable flock, i.e, $\mathcal{A} < \mathcal{A}_c$, the \SR{dynamics implied by our theory on long time-scales} is equivalent to the Toner-Tu equations, with the coefficient of the advective term shifted by a contribution proportional to $\mathcal{A}$. Thus\SR{, through the connection to an explicitly nonequilibrium aligning interaction, our results also provide a physically clear and appealing demonstration of the role of broken detailed balance in dictating distinct advection speeds for the density and the polarisation or velocity fields}. 
\SR{The antisymmetric coupling} 
thus plays a dual role, providing a natural mechanism for a difference between the \SR{speeds of information transfer about orientation and density} \textit{and}, if large enough, destabilizing the flock. \AMR{The role of the antisymmetric coupling in leading to different speeds of orientation and density advection is best understood in a system \emph{without} a motility -- polar spins on a lattice -- in which the sole presence of non-mutual torques leads to the \emph{self-advection} of the polarisation field without any mass motion, a well-known feature of flocking models normally associated with motility \cite{RMP}.}

We now discuss how we obtain these results. In Sec \ref{Micro_model} we arrive at the continuum equations for a flock with \SR{asymmetric aligning interactions} by starting with a microscopic model in which it is an explicit ingredient. \AMR{We also demonstrate that a microscopic model of \emph{immotile} spins on a lattice interacting via such interactions has a continuum equation for the magnetisation that is equivalent to the velocity equation of a flock in which the number of flockers is not conserved.} 
In Sec \ref{Macro_mod} we consider the dynamics of a system of motile polar rods on a substrate, \SR{\it without explicit microscopic antisymmetric interactions} \cite{Harsh}, retaining the dynamics of the spin angular momentum and demonstrate that motility induces \SR{such an effective antisymmetry} in the continuum equations. Then in Sec. \ref{stab_an}, we discuss the linear stability of a flock with \SR{an antisymmetric contribution to the aligning torque}. We conclude with a summary of our findings in \ref{concl}.

\section{Microscopic model}
\label{Micro_model}
Consider a flock in the $xy$ plane. Let the $i-$th bird have velocity ${\bf  
v}_{\alpha} \equiv v_0 \hat{\bf v}_{\alpha}$ with fixed magnitude $v_0$ and 
classical ``spin'' angular momentum ${\bf 
s}_{\alpha}$, about its centre of mass, along $\hat{\bf z}$. Let us describe the aligning 
interaction of neighbouring birds as a torque  
\beq
\label{torque}
\dot{\bf s}_{i} = 
\sum_{j}J_{ij}\hat{\bf v}_{i} \times \hat{\bf v}_{j}
\eeq
due to birds $j$ neighbouring $i$, which rotates the direction of the velocities:  
\beq
\label{rotation}
\dot{\bf v}_{i} = {{\bf s}_{i} \over \chi} \times {\bf v}_{i}, 
\eeq
where $\chi$ is a rotational inertia \cite{ISM,silent}. As remarked above we must allow for processes 
that do not conserve angular momentum \cite{air}. Specifically, we must 
allow for the possibility that the 
coupling $J_{ij}$ is non-symmetric. The result is that the rate at 
which bird $i$ turns to align with $j$ differs from that at which 
$j$ turns to align with $i$. Such \SR{non-reciprocal torques are analogous to an antisymmetric exchange coupling \cite{jayajit,DM}. They }violate angular momentum conservation while preserving rotation 
invariance, because the aligning field does not arise from an energy 
function \SR{\cite{behav} and} is under no obligation to do so. Making 
the physically reasonable approximation that the interactions of birds are 
left-right symmetric implies that $J_{ij}$ should have an antisymmetric 
part when birds $i$ and $j$ are one behind the other -- possibly 
through polar asymmetries of airflow or vision -- and not when they are 
side-by-side. Of course, even the symmetric part of $J_{ij}$ 
should in general be anisotropic and should thus differ for longitudinal and lateral 
neighbours. 
However, this latter asymmetry is not of much consequence for the issues considered here; at least within a linearized theory it can be removed by anisotropic rescaling of coordinates.   

With these considerations, the interactions the total change of the spin angular momentum of a bird $i$, interacting with all other birds $j$ \SR{in a neighbourhood 
$\mathcal{R}_i$} around it is 
\begin{equation}
\label{particleeq}
\dot{{\bf s}}_i+\frac{\eta}{\chi}{\bf s}_i=\SR{\sum_{{\bf r}_j \in \mathcal{R}_i}}[\tilde{J}+\tilde{A}(\hat{v}_i\cdot\hat{r}_{ij})]\hat{v}_i\times\hat{v}_j
\end{equation}
where $\eta$ is the friction with the ambient medium. We examine the dynamics of two particles interacting via this \SR{asymmetric} interaction in the appendix \ref{Two_part}. \SR{One distinctive feature of the two-particle dynamics of \eqref{torque} and \eqref{rotation}, when compared with the case of symmetric interactions, can however be drawn immediately: if the asymmetry is so large that $\bar{A}>\bar{J}$,} particles tend to align with those in from of them while tending to {\it anti-align} with those directly behind them. Therefore, two back-to-back particles will experience a torque which will {\it reinforce} their anti-alignment, in contrast to a model with a purely symmetric \SR{interaction}. However, it could be argued that this distinction, while important in a model {\it without} motility, may be less consequential in a motile system -- two back-to-back particles would move in opposite directions out of each others interaction range and therefore, would stop interacting almost immediately -- with a particle interacting longer with those in front of it than those behind it, on average, purely due to motility. Therefore, to examine whether \SR{asymmetric aligning torques have} any qualitative effect on the dynamics of a flock, we now construct the continuum equations of motion for this system.

We define the \SR{density} $\rho({\bf x},t)=\sum_i\delta({\bf x}-{\bf x}_i)$, the spin angular momentum density ${\bf s}({\bf x},t)=\sum_i{\bf s}_i\delta({\bf x}-{\bf x}_i)$, and the velocity (not the momentum density) or \SR{equivalently} the local polarisation as
\begin{equation}
{\bf v}({\bf x},t)=\frac{\sum_i{\bf v}_i\delta({\bf x}-{\bf x}_i)}{\sum_i\delta({\bf x}-{\bf x}_i)}.
\end{equation}
\SR{A favoured} mean speed $v_0$ in the coarse-grained model is implemented through a potential 
\begin{equation}
\label{poteneq}
\SR{U={1 \over 2}\int_{\bf x}\left[-{{\bf v}\cdot{\bf v}}+{1 \over 2 v_0^2}{({\bf v}\cdot{\bf v})^2}+{K(\nabla {\bf v})^2} \right]}
\end{equation}
where we have also additionally included an elastic term that penalises gradients in ${\bf v}$.
In terms of these coarse-grained variables, the equations of motion have the form
\begin{equation}
\label{conteq}     
D_t\rho=-\rho\nabla\cdot \SR{\bf v}
\end{equation}
\begin{equation}
\label{velpde}
D_t{\bf v}=\frac{1}{\chi}{\bf s}\times{\bf v}-\frac{1}{\rho}\nabla f(\rho)-\Gamma_v\frac{\delta U}{\delta{\bf v}}
\end{equation}
where $f(\rho)$ is a function of the density and $\Gamma_v$ is a kinetic coefficient,
\begin{equation}
\label{spinpde}
D_t{\bf s}=\frac{\mathcal{A}}{v_0^3}{\bf v}\times({\bf v}\cdot\nabla{\bf v})+\frac{J}{v_0^2}{\bf v}\times\nabla^2{\bf v}+\frac{J_A}{v_0^4}{\bf v}\times[({\bf v}\cdot\nabla)^2{\bf v}]-\frac{\eta}{\chi}{\bf s}
\end{equation}
where $D_t = 
\partial_t + \mathbf{v} \cdot \nabla$ is the material derivative, without the possibility at this stage of an arbitrary advection coefficient, and the antisymmetric coupling $\mathcal{A}$ acts between fore-aft neighbours as defined \SR{\it locally} by ${\bf v}$. In going from \eqref{particleeq}, which has the form of a {\it difference} equation to the continuum differential equation for the coarse-grained spin angular momentum \eqref{spinpde}, we have implicitly \SR{assumed that the interaction neighbourhood of a bird is local and based on distance, and thus} introduced a mean interbird spacing $a\sim\rho_0^{-1/d}$, where $\rho_0$ is the mean density and $d$ is the dimensionality, in terms of which $J = a^2\tilde{J}\sim\rho_0^{-2/d}\tilde{J}$ and $\mathcal{A} = 2a\tilde{A}\sim 2\rho_0^{-1/d}\tilde{A}$. The ratio $J/\mathcal{A}$ thus has units of length. Equations (\ref{velpde}) and (\ref{conteq}) are as in the original inertial spin model \cite{ISM}, but, crucially, (\ref{spinpde}) has a qualitatively new term $\mathcal{A} 
\hat{\bf v}\times\partial_\parallel \hat{\bf v}$ from the antisymmetric \SR{coupling}, where $\partial_{\parallel} = \hat{\bf v} \cdot \nabla$. We have also introduced a term $\propto J_A$ which arises from an allowed {\it anisotropic} contribution to the symmetric \SR{coupling} $J$. In equilibrium, $\mathcal{A}$ has to be $0$ and both $J$ and $J_A$ \SR{have} to arise from the potential in \eqref{poteneq}; i.e. $J|_{\text{eq}}=K$ and $J_A|_{\text{eq}}=0$ for the potential defined by \eqref{poteneq}. To obtain an anisotropic contribution to the \SR{interaction}, one has to introduce an anisotropic energy cost for gradients in ${\bf v}$ in equilibrium. However, in the nonequilibrium model we consider here, $J$ and $J_A$ should be considered to be arbitrary coefficients which are not obliged to have any relation with $K$, the energy cost for gradients in ${\bf v}$. 

\AMR{Note that a spin dynamics of the form \eqref{particleeq} requires \emph{polarity} but \emph{not motility}. To see this more clearly, take $\hat{v}_i$ to be the direction of the $X-Y$ spin permanently assigned to site $i$ on a lattice with $\hat{r}_{ij}$ being one of the lattice vectors and the summation over $\mathcal{R}_i$ denoting a sum over nearest neighbours. Then, the coarse-grained variable ${\bf v}({\bf x}, t)$ has the interpretation of the polarisation of this spin system. The equation of motion for ${\bf v}$, replacing \eqref{velpde}, in this case is 
\begin{equation}
\label{magpde}
\partial_t{\bf v}=\frac{1}{\chi}{\bf s}\times{\bf v}-\Gamma_v\frac{\delta U}{\delta{\bf v}}.
\end{equation}
Since there is no mass motion and density variation due to the spins being associated with lattice points, the material derivative is replaced by a simple time derivative and the gradient of density in \eqref{velpde} is absent. Similarly, there is no density equation and the equation of motion for the spin angular momentum is simply \eqref{spinpde} with $D_t$ replaced by $\partial_t$ \cite{comm_diff}. However, while the equations of motion are similar, we emphasise that the physical situation that this set of equations model is very different from the one described by \eqref{velpde}, \eqref{spinpde} and \eqref{conteq}; the latter deals with the dynamics of a motile flock with asymmetric interaction while this with a static spins on a lattice which exert non-mutual torques on each other.
}

\section{Adsorbed active polar liquid crystals with spin}
\label{Macro_mod}
\noindent\SR{In the last section we considered a model in which individual birds have explicitly asymmetric interactions. However,} the continuum equations that we arrived at also describe the \SR{collective} dynamics of \SR{aligning} motile \AMR{polar} rods on a substrate, such as those in \cite{Harsh} \AMR{or multiple other systems described using the Vicsek model} in which the microscopic \AMR{polar, aligning} interaction is purely due to contact mechanics and, therefore, symmetric \cite{comm3}. In other words, \SR{the breaking of time-reversal invariance through motility \cite{LPDJSTAT} generates an  asymmetric effective interaction {\it even when} the microscopic interaction of the constituents is purely symmetric as it arises from a pair potential}. To see this, we start with a theory of polar liquid crystal hydrodynamics \cite{kung}  modified to include the dynamics of spin angular momentum \cite{Stark} and a momentum sink. We discuss \SR{this} theory in detail in appendix \ref{Pol_rods}, and point out \SR{in the main text how an \SR{antisymmetric contribution} effectively emerges due to the motility from this description.}
We construct the coupled hydrodynamic equations for \SR{\textit{separate}} velocity and orientation fields $\mathbf{u}$ and $\mathbf{v}$ and spin angular momentum density field $\mathbf{s}$, together with continuity $\partial_t \rho + \nabla \cdot (\rho \mathbf{u}) = 0$ for the density field $\rho$.
\begin{equation}
\label{ppde}
D_t\mathbf{v}=\frac{1}{\chi}\mathbf{s}\times\mathbf{v}-\Gamma_v\mathbf{h}-\frac{\Gamma'_u}{|v|^2}\mathbf{v}(\mathbf{v}\cdot\mathbf{u})
\end{equation}
\begin{multline}
\label{spde}
D_t\mathbf{s}=-\mathbf{v}\times\mathbf{h}+\Gamma_\omega\boldsymbol{\omega}-\frac{\eta}{\chi}\mathbf{s}-\Gamma_D\mathbf{v}\times(\mathbf{v}\cdot\bsf{D})\\+\Gamma_D'{\bf v}\times(\nabla\cdot\bsf{D})-\Gamma_u(\mathbf{v}\times\mathbf{u}),
\end{multline}
where \SR{$\bsf{D} =[\nabla \mathbf{u} + (\nabla \mathbf{u})^T]/2$} is the symmetric part of the velocity gradient, $\boldsymbol{\omega}=(\nabla\times\mathbf{u})/2$ \SR{is the vorticity, $\Gamma_\omega$ and $\eta/\chi$ govern the relaxation of the spin angular momentum to the fluid vorticity and the damping of $s$ by the substrate,} and the term with $\Gamma_u$ denotes the torque exerted on the polarisation by an imposed flow. The equation of motion for the velocity field is 
\begin{equation}
\label{vpde}
\rho(\partial_t + \mathbf{u} \cdot \nabla)\mathbf{u} + \Gamma \mathbf{u} =\zeta \mathbf{v} + ....
\end{equation}
where $\Gamma$ in (\ref{vpde}) denotes damping by a substrate and the ellipsis denotes \SR{terms arising from pressure, viscosity, and order-parameter stresses}. Eq \eqref{ppde} and \eqref{spde} are exactly the equations that one would have obtained for a uniaxial \SR{polar} liquid crystal on a substrate \SR{with vector orientational order parameter $\mathbf{v}$}. That the system is intrinsically nonequilibrium enters only at one point: the forcing $\zeta \mathbf{v}$ in \eqref{vpde} for the velocity and the polarisation. 
If one integrated out ${\bf s}$, by taking its dynamics to be fast, the two terms with the coefficients $\Gamma_u$ and $\Gamma'_u$ would have yielded the \SR{``weathercock'' term \cite{Harsh} familiar from theories of polar liquid crystals on substrates \cite{Ano_pol, LPDJSTAT, Lauga}}. Further, $\Gamma_\omega$ and $\Gamma_D$ together yield the flow alignment term familiar from theories of nematic liquid crystal \cite{Stark, deGen, flowalign} in this limit. If we now take the opposite limit, which is relevant for inertial spin models, and eliminate ${\bf u}$ in favour of ${\bf v}$, ${\bf u}\sim(\zeta/\Gamma){\bf v}$, it is immediately obvious that a term of the form ${\bf v}\times({\bf v}\cdot\nabla{\bf v})$ in the $D_t{\bf s}$ equation emerges via the term $\propto \Gamma_D$ in \eqref{spde}. This has exactly the form of the \SR{antisymmetric coupling} discussed above, though no microscopic antisymmetric interaction is implied in this case. Furthermore, it is clear that extra active contributions to the \SR{symmetric coupling} also arise from the $\Gamma_D'$ term, when ${\bf u}$ is replaced by $(\zeta/\Gamma){\bf v}$.
This emergence of an effective \SR{asymmetric interaction} due to motility can be rationalised as follows: since the particles move in the direction they are pointing in, they {\it move towards} the particles in front of them and {\it away from} those that are behind them. Since the interaction range is finite, the particle on average therefore interacts for a longer time with the particles in \SR{front} of it. Therefore, when the particles motion is averaged over a coarse-graining time, an effective \SR{asymmetric} interaction should emerge (in the sense it should on average be more affected by particles in front of it than behind it) since on average, particles in front of it should remain in \SR{its} neighbourhood for longer.

\section{Stability analysis}
\label{stab_an}
The \SR{wavevector-independent} damping $\propto\eta$ in \eqref{spinpde} or the equation obtained by replacing ${\bf u}$ by $\zeta/\Gamma{\bf v}$ in \eqref{spde} implies that $\mathbf{s}$ is a fast variable, relaxing on non-hydrodynamic timescales to a value determined by $\mathbf{v}$ and $\rho$. \AMN{Therefore, our model which has two true hydrodynamic quantities -- transverse fluctuations of broken rotational symmetry and conserved density -- must belong to the Toner-Tu universality class. In other words, upon eliminating ${\bf s}$, one should obtain the equations of motion first discussed in \cite{TT_PRL}.} \AMN{We use \eqref{velpde} and \eqref{spinpde} to demonstrate this explicitly. A formal solution of ${\bf s}$ in terms of a function of ${\bf v}$ can be obtained from \eqref{spinpde} as
\begin{multline}
{\bf s}=\frac{\chi}{\eta}\left[1+\sum^\infty_{n=1}\left(\frac{-\chi}{\eta}\right)^nD^n_t\right]\bigg[\frac{\mathcal{A}}{v_0^3}{\bf v}\times({\bf v}\cdot\nabla{\bf v})\\+\frac{J}{v_0^2}{\bf v}\times\nabla^2{\bf v}+\frac{J_A}{v_0^4}{\bf v}\times[({\bf v}\cdot\nabla)^2{\bf v}]\bigg]
\end{multline}
We insert this solution in \eqref{velpde} and expand in $1/\eta$. We calculate the dynamics of the component of the velocity field transverse to the mean motion $\delta v_\perp$, where $\perp$ and $\parallel$ denote the directions perpendicular to and along the mean motion, and only retain linear terms with up to two gradients and nonlinear terms with two fields and one gradient (see Appendix \ref{ElimSpin} for details), as is traditional in the Toner-Tu model
\begin{multline}
\label{TonTulike}
\partial_t \delta v_\perp+\left(1-\frac{\mathcal{A}}{v_0\eta}\right){\bf v}\cdot\nabla \delta v_\perp=-\frac{1}{\rho}\partial_\perp f(\rho)+\frac{J}{\eta}\nabla^2 \delta v_\perp\\+\Gamma_vK\nabla^2 \delta v_\perp+\left(\frac{J_A}{\eta}-\frac{\chi\mathcal{A}^2}{\eta^3}\right)\partial_\parallel^2 \delta v_\perp+\frac{2\chi\mathcal{A}\sigma}{\eta^2}\partial_\parallel\partial_\perp\delta\rho
\end{multline}
where we have written $({\bf v}\cdot\nabla)^2$ as $v_0^2\partial_\parallel^2$ since the other nonlinear terms originating from it are subdominant to the advective nonlinearity on the L.H.S. Here, $\sigma=f'(\rho_0)/\rho_0$ with the prime denoting differentiation with $\rho$ and $\rho_0$ being the steady state density and $\rho=\rho_0+\delta\rho$. While this equation is formally obtained via a $1/\eta$ expansion, it is correct to all orders in $1/\eta$ at this order in gradients and fields. It can be easily checked that higher order in $1/\eta$ terms to do not modify the equation for $D_t v_\perp$ at this order in gradients and fields.
}
\AMN{The continuity equation (\ref{conteq}) remains unchanged.   }
\AMN{We thus recover the equation of motion for the transverse fluctuations that are obtained from the Toner-Tu \cite{tonertu} model}. Interestingly, the antisymmetry parameter $\mathcal{A}$ provides a natural mechanism for a coefficient different from unity for the advection term in \eqref{TonTulike}, a feature of the Toner-Tu formulation \cite{tonertu}, \SR{often simply} ascribed to the lack of Galilean invariance in the model. However, losing Galilean invariance alone cannot generate such a term: the absence of detailed  balance, as signalled by $\mathcal{A}$, is crucial \cite{LPDJSTAT}. In an \textit{equilibrium} fluid moving in contact with a momentum sink the coefficients in the momentum and density equations {\it must} agree. Note that $\mathcal{A}$ leads to a fore-aft asymmetry in the local transfer of orientational information. It is therefore natural that it should give rise to a disturbance speed different from the flock speed. \AMR{This becomes clearer in the spin system on a lattice discussed at the end of Sec. \ref{Micro_model} (see \eqref{magpde}) for which the expression for the coarse-grained spin polarisation is similar to the velocity equation \eqref{TonTulike}:
\begin{multline}
\label{MalTonTulike}
\partial_t \delta v_\perp-\frac{\mathcal{A}}{v_0\eta}{\bf v}\cdot\nabla \delta v_\perp=\frac{J}{\eta}\nabla^2 \delta v_\perp\\+\Gamma_vK\nabla^2 \delta v_\perp+\left(\frac{J_A}{\eta}-\frac{\chi\mathcal{A}^2}{\eta^3}\right)\partial_\parallel^2 \delta v_\perp
\end{multline}
Note however, that there is no dependence on density in \eqref{MalTonTulike} due to the assumption that the spins are on a lattice. This is therefore equivalent to the equation used for the \emph{velocity} field in Malthusian Toner-Tu \cite{tonertu_Malthus} albeit here realised not in a motile system but on a lattice-based system. We have therefore successfully constructed a on-lattice variant of the usual flocking model, which may significantly simplify its simulation. Furthermore, the presence of an advection term, \emph{even in this immotile system}, implies that this model of spins on a lattice must have long-range order at low noise-strengths when all the diffusive terms in \eqref{MalTonTulike} are stabilising.
}

\AMN{Both \eqref{TonTulike} and \eqref{MalTonTulike} demonstrate that \SR{beyond a critical value of $\mathcal{A}$} 
\begin{equation}
\label{instabcond}
\mathcal{A}_c \equiv \eta \sqrt{\bar{J} / \chi},
\end{equation}
where $\bar{J}=J+J_A+\Gamma_vK\eta$, the effective diffusivity for longitudinal perturbations \SR{turns} \emph{negative} signalling a buckling instability of the ordered phase. Such a long-wavelength instability implies that one \SR{must include terms at higher order in wavenumber to} stabilise the system at small scales. Such higher order terms in the $\delta v_\perp$ equation, upon eliminating the  spin variable, can be obtained using the method described above. However, it is quite cumbersome in practice. Instead, we examine the full mode structure (i.e., at arbitrary wavenumber) implied by the equations \eqref{conteq}, \eqref{velpde} and \eqref{spinpde} and from that obtain the different scaling regimes of the relaxation rate as a function of wavenumber including, finally, the wavenumbers above which the system is stable.}

\begin{figure}
  \includegraphics[width=8 truecm]{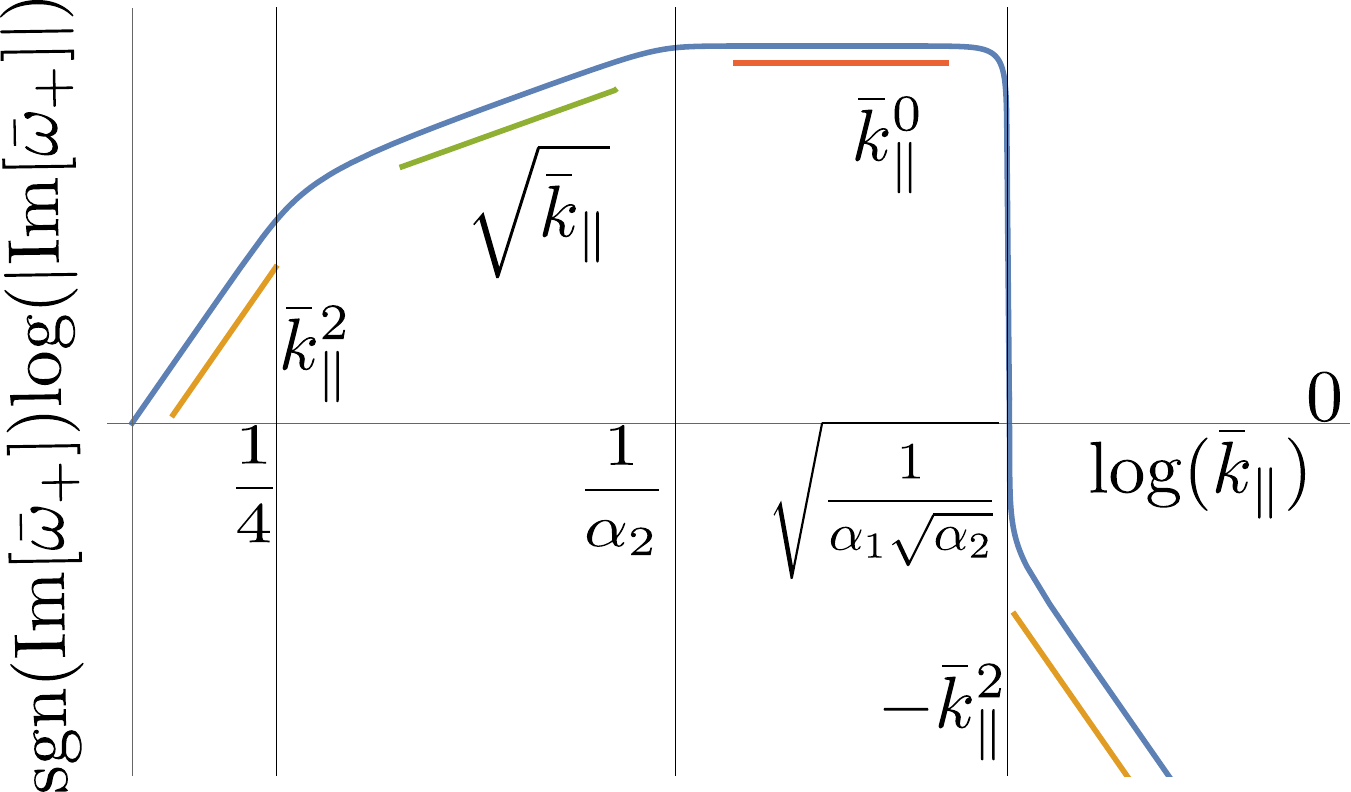}
  \caption{\SR{Log-log plot of the absolute value of the} imaginary part of the eigenfrequency $\bar{\omega}_+$ in \eqref{mode_struct_full} showing the different scaling regimes. \SR{To distinguish stable and unstable regimes we multiply the ordinate by $\text{sgn}(\text{Im}[\bar{\omega}_+])$.}}
  \label{fig:md_st}
\end{figure}

\AMN{To this end we linearise \eqref{conteq}, \eqref{velpde} and \eqref{spinpde}  about an ordered steadily moving} state: $\mathbf{v}={\bf v}_0 +\delta\mathbf{v}$, $\rho=\rho_0 
+\delta\rho$, ${\bf s} = {\bf 0} + s\hat{\bf z}$. Working in a frame moving with the mean 
velocity of flock
        \begin{equation}
        \label{vperpeq}
        \partial_t \delta v_\bot = -\sigma\partial_\bot \delta\rho + 
\frac{v_0}{\chi}s+\Gamma_v K\nabla^2\delta v_\perp
        \end{equation}
        \begin{equation}
\label{spinlineq}
        \partial_t s = {J\over v_0}\nabla^2\delta v_\perp+{J_A\over v_0}\partial_\parallel^2\delta v_\perp + {\mathcal{A} 
\over v_0} \partial_\parallel\delta v_\perp - \frac{\eta}{\chi}s
        \end{equation}
      \begin{equation}
\label{denslineq1}
      \partial_t\delta\rho = -\rho_0\partial_\perp\delta v_\perp,
      \end{equation}
We first consider the dynamics in the absence of density fluctuations, which is \SR{technically acceptable in flocks 
} in which birth and death keep the density \SR{fixed on average without a strict} conservation law \AMR{and, as discussed earlier, also models spins on a lattice interacting via asymmetric interaction \cite{comm2},} and discuss the mode structure including the density fluctuations in appendix \ref{cons_dens}. \AMN{To first check that our elimination of the spin variable presented above was consistent, we calculate the 
the eigenfrequencies implied by \eqref{vperpeq} and \eqref{spinlineq} in the limit of small wavevectors, i.e., in the Toner-Tu regime
\begin{equation}
\omega_1=-i\frac{\eta}{\chi}+\frac{\mathcal{A}}{\eta}k_\parallel
\end{equation}
\begin{equation}
\label{mode2}
\omega_2=-\frac{\mathcal{A}}{\eta}k_\parallel-\frac{i}{\eta}\left[(J+\Gamma_v K\eta) k^2+J_Ak_\parallel^2-\frac{\mathcal{A}^2\chi}{\eta^2}k_\parallel^2\right]
\end{equation}
$\omega_1$ corresponds to the fast decay of $s$, at a rate $\eta/\chi$ in the limit of zero wavenumber. Unsurprisingly, $\omega_2$ implies a buckling instability whose effect is largest for the disturbances with wavevectors aligned precisely along the ordering direction since the coupling of orientation to spin angular momentum through $\mathcal{A}$ 
enters only the modes with $k_{\Vert} \neq 0$. For $\mathcal{A} > 0$, which corresponds to the case where a polar spin responds more to the one ahead of it than to the one behind, we see from \eqref{mode2} that the unstable disturbance travels towards the rear of the ordered phase, which is physically reasonable. } 


To examine the behaviour of at high wavevectors, we specialise to distortions with wavevectors purely along $k_\parallel$, nondimensionalise $\bar{\omega}\equiv\omega\chi/\eta$ and $\bar{k}_\parallel\equiv(\mathcal{A}\chi/\eta^2)k_\parallel$ and define the nondimensional quantities $\alpha_1=\Gamma_v K\eta^3/\mathcal{A}^2\chi$ and $\alpha_2=({J+J_A})\eta^2/\mathcal{A}^2\chi$ to obtain the eigenfrequencies
\begin{equation}
\bar{\omega}_\pm=\frac{1}{2}\left[-i(1+\alpha_1\bar{k}^2_\parallel)\pm\sqrt{-(\alpha_1\bar{k}_\parallel^2-1)^2+4\left(-i\bar{k}_\parallel+\alpha_2\bar{k}_\parallel^2\right)}\right].
\label{mode_struct_full}
\end{equation}
Now, examining the  eigenvalue $\bar{\omega}_+$, it is easy to see that at small $\bar{k}_\parallel$, it goes as $-\bar{k}_\parallel-i(\alpha_1+\alpha_2-1)\bar{k}^2_\parallel$, i.e., has a positive growth rate for $\alpha_1+\alpha_2<1$ which is simply the condition $\mathcal{A}>\mathcal{A}_c$ while at large $\bar{k}_\parallel\gg1$ it goes as $-i\alpha_1\bar{k}_\parallel^2+i\alpha_2/\alpha_1$ i.e., it decays as $-\bar{k}_\parallel^2$. $\omega_-$ is always stabilising and decays as $-i$ at small $\bar{k}_\parallel$ and as $-i(\alpha_1+\alpha_2)/\alpha_1$ at large $\bar{k}_\parallel$. However, the crossover of $\text{Im}[\omega_+]$ from a growth rate $\propto \bar{k}_\parallel^2$ at small wavevectors to a {\it decay} rate $\propto \bar{k}_\parallel^2$ at large wavevectors passes through two intermediate scales. First, for $1/\alpha_1, 1/\alpha_2\gg \bar{k}_\parallel\gg 1/4$, $\text{Im}[\bar{\omega}_+]$ crosses over from a growth rate of $\bar{k}^2_\parallel$ to $\sqrt{\bar{k}_\parallel}$, then (assuming $\alpha_2>\alpha_1$) for $\bar{k}_\parallel>1/\alpha_2$, it becomes independent of wavevector before finally decaying as $-\bar{k}^2_\parallel$ for $\bar{k}^2_\parallel>1/\alpha_1\sqrt{\alpha_2}$ (see Fig. \ref{fig:md_st}).
Somewhat unusually, the mode structure displayed in Fig. \ref{fig:md_st} and implied by \eqref{mode_struct_full} passes through a $\bar{k}^0_\parallel$ plateau before decaying. \SR{Thus, as we said in the Introduction, there is a broad band of growing modes, not a sharply defined wavenumber of fastest growth. Possibly therefore, 
the emergent patterns beyond the linear instability will have a broad range of length scales }
\AMN{unlike in multiple other active systems with a diffusive instability \cite{Ano_apol, Aran_PRX, Dunkel}}. Further, it is clear from \eqref{vperpeq}-\eqref{denslineq1}  that the density fluctuations are unaffected by the orientation and spin fields for distortions with wavevector $k_\parallel$ and therefore, the eigenfrequncies $\bar{\omega}_\pm$ for perturbations purely along the ordering direction remain the same even for systems with a fixed number of particles.

Finally, can we say something about the nonlinear dynamics of this system? Within a \SR{``Malthusian'' \cite{tonertu_Malthus}} approximation (i.e., \SR{where the density has no conservation law but is constant in the mean}), the lowest order nonlinearity in \eqref{vperpeq} and \eqref{spinlineq} arises from the antisymmetric \SR{coupling}: $\SR{(\mathcal{A}/v_0^2)}\delta v_\perp\partial_\perp\delta v_\perp$ in the $\dot{s}$ equation. This corresponds to the usual advective nonlinearity in the Toner-Tu theory \cite{tonertu}. However, notice that the flock is unstable for wavevectors along the direction of motion i.e., the effective longitudinal diffusivity in the Toner-Tu regime turns negative. There can be no graphical correction to the longitudinal diffusivity due to the advective nonlinearity which vanishes for wavevectors along the ordering direction. This would seem to imply that a nonlinear stabilisation of the ordered phase at large enough scales even when the linear theory predicts an instability as in \cite{Rayan}, via a change in sign of the effective longitudinal diffusivity upon averaging over small scales \cite{Rahul_KS}, is ruled out here. This is possibly the case. However, we note that (i) the fact that nonlinearities cannot renormalise the longitudinal diffusivity is only correct in the Malthusian limit in which concentration is taken to be locally fixed. (ii) Even for Malthusian flocks, higher order nonlinear terms such as $\partial_\perp\delta v_\perp\partial_\parallel\delta v_\perp$, which arises from a torque $\propto {\bf v}\times[(\nabla\cdot{\bf v}){\bf v}\cdot\nabla{\bf v}]$, that are less relevant than the advective nonlinearity, may nevertheless \SR{modify the} effective longitudinal diffusivity upon coarse-graining. We cannot therefore conclusively rule out a nonlinear stabilisation in active models with \SR{asymmetric} \SR{interactions}. However, our preliminary simulations on microscopic spins \AMR{on lattices} \SR{interacting} via \eqref{torque} and \eqref{rotation} \SR{suggest} that this is not the case -- instead, \AMR{beyond the threshold of the instability a family of inhomogeneous states appear, transitioning to  a statistically isotropic and homogeneous state at higher values of the asymmetry. We are now working on establishing the stability of these inhomogeneous states as well as characterising the transition from them to the statistically isotropic phase.} 

\section{Conclusions}
\label{concl}
In this paper we demonstrated that the hallmark of flocking models -- distinct speeds of \SR{alignment} information and \SR{density disturbances} -- \SR{is usefully viewed as arising from} a fundamentally nonequilibrium antisymmetric \SR{interaction between spins}, either microscopically introduced or emergent, 
in a theory that explicitly retains the non-hydrodynamic spin angular momentum field. \AMR{This is most clearly evident in a model of on-lattice spins exerting non-mutual torques on each other, in which mass motion is forbidden, but alignment interaction is advected at a non-zero speed, resulting in long-range order.}
In conjunction with \cite{LPDJSTAT} this firmly establishes that this feature of flocking models require {\it both} the breaking of Galilean invariance {\it and} time-reversal symmetry. We further established that a flock suffers a small wavenumber buckling instability for sufficiently high values of this antisymmetric \SR{interaction}. Our theory should be testable in simulations of \SR{microscopic models with inertial spin dynamics featuring explicitly asymmetric aligning} interactions \cite{Chate_asym} or in experiments and mechanically faithful simulations of \SR{flocking in motile} polar rods on substrates \cite{Harsh} where, in principle, the antisymmetric \SR{coupling} coefficient can be measured. \SR{As the instability sets in at long enough wavelengths, the message is that a strong fore-aft asymmetry in sensing the orientation of neighbours, together with a significant inertial lag in responding, is disadvantageous to the formation of large flocks. Apart from serving as one more post facto explanation for limits on group size, our instability mechanism should be borne in mind when designing robotic swarms.}


\begin{acknowledgments}
SR was supported by a J C Bose Fellowship of the SERB (India) and the Tata Education and Development Trust. SR also thanks the Simons Foundation and the KITP for support. SR is Adjunct Professor at TIFR Hyderabad, where a part of this work was done. 
\end{acknowledgments}

\onecolumngrid
\appendix
\section{Two particle dynamics}
\label{Two_part}
Consider two spins fixed in space. The angular dynamics of the two spins are described by 
\begin{equation}
\dot{s}_1+\frac{\eta}{\chi}s_1=J{\bf v}_1\times{\bf v}_2+A({\bf v}_1\cdot\hat{r}_{12}){\bf v}_1\times{\bf v}_2
\end{equation}
and 
\begin{equation}
\dot{s}_2+\frac{\eta}{\chi}s_2=J{\bf v}_2\times{\bf v}_1+A({\bf v}_2\cdot\hat{r}_{21}){\bf v}_2\times{\bf v}_1
\end{equation}
where $\hat{r}_{12}=-\hat{r}_{21}$. Taking the spins ${\bf v}_1=(\cos\theta_1,\sin\theta_1)$ and ${\bf v}_2=(\cos\theta_2,\sin\theta_2)$, where $\theta_1$ and $\theta_2$ are both measured from the line joining the first spin to the second (see Fig \ref{twopart}), these equations become
\begin{equation}
\chi\ddot{\theta}_1+{\eta}\dot{\theta}_1=-J\sin(\theta_1-\theta_2)-A\cos\theta_1\sin(\theta_1-\theta_2)
\end{equation}
and
\begin{equation}
\chi\ddot{\theta}_2+{\eta}\dot{\theta}_2=J\sin(\theta_1-\theta_2)-A\cos\theta_2\sin(\theta_1-\theta_2)
\end{equation}
Defining $\bar{\theta}=\theta_1+\theta_2$ and $\delta\theta=\theta_1-\theta_2$. Then using the trigonometric identities $\cos\theta_1+\cos\theta_2=2\cos[(\theta_1+\theta_2)/2]\cos[(\theta_1-\theta_2)/2]$ and $\sin\theta_1+\cos\theta_2=-2\sin[(\theta_1+\theta_2)/2]\cos[(\theta_1-\theta_2)/2]$, we obtain
\begin{equation}
\label{sum}
\chi\ddot{\bar{\theta}}+{\eta}\dot{\bar{\theta}}=-2A\sin(\Delta\theta)\cos\left(\frac{\bar{\theta}}{2}\right)\cos\left(\frac{\Delta\theta}{2}\right)
\end{equation}
\begin{equation}
\label{diff}
\chi\Delta\ddot{\theta}+{\eta}\Delta\dot{{\theta}}=-2\sin(\Delta\theta)\left[J-A\sin\left(\frac{\bar{\theta}}{2}\right)\sin\left(\frac{\Delta\theta}{2}\right)\right]
\end{equation}
The obvious and trivial static solution is $\Delta\theta=0$ which is stable. The other static solutions are for $\Delta\theta=\pi$ and $\Delta\theta=-\pi$, i.e., spins that point away from each other. For $\pi$, to first order in perturbation $\Delta\theta=\pi+\epsilon$, the linearised, overdamped equations of motion for $\epsilon$ is 
\begin{equation}
\eta\dot{\epsilon}=2\epsilon\left[J-A\sin\left(\frac{\bar{\theta}}{2}\right)\right]
\end{equation}
and for $-\pi$, it is 
\begin{equation}
\eta\dot{\epsilon}=2\epsilon\left[J+A\sin\left(\frac{\bar{\theta}}{2}\right)\right].
\end{equation}
Thus, if $A\sin\left({\bar{\theta}}/{2}\right)>J$, deviations of $\Delta\theta$ away from $\pi$ decay and if $A\sin\left({\bar{\theta}}/{2}\right)<0$ and $|A\sin\left({\bar{\theta}}/{2}\right)|>J$ deviations of $\Delta\theta$ away from $-\pi$ decay. These situations corresponds to the following: take $|A|>J$, in which case, spins try to anti-align with the ones behind them. Then, two spins that are back to back with each other both have their partners behind them. This situation is stable.
\begin{figure}
 \includegraphics[width=7cm]{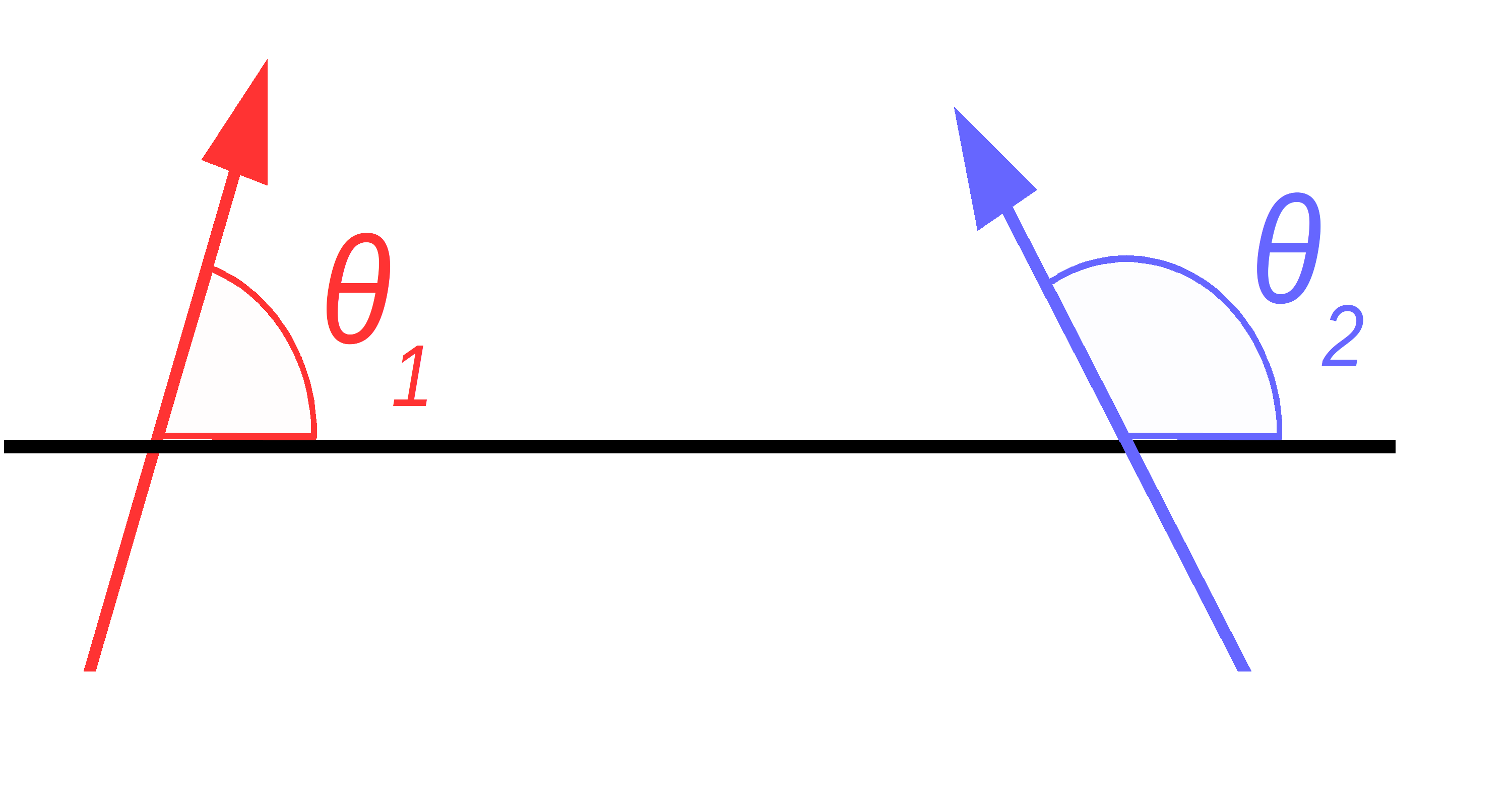}
\caption{Two fixed spins with the angle measured from the line joining the red spin to the blue spin.} 
\label{twopart}
\end{figure}
It is easy to check that there is no situation in which $\Delta\theta$ remains fixed, i.e., the R.H.S. of \eqref{diff} becomes $0$ but $\bar{\theta}$ doesn't with the R.H.S. of \eqref{sum} becomes a constant, which  would mean that the two spins rotate in concert, chasing each other forever, in an overdamped system (ignoring the inertial term). Such a solution would require the term in the square bracket in \eqref{diff} to vanish (since $\sin\Delta\theta$ would be a non-zero constant) for all $\bar{\theta}$. However, when $J<A$ another fixed points at $\bar{\theta}=\pi$, $\sin(\Delta\theta_0/2)=J/A$ is possible. We check the stability of this fixed point by taking $\bar{\theta}=\pi+\delta$ and $\Delta\theta=\Delta\theta_0+\epsilon$. The linearised, overdamped equations of motion in this case are 
\begin{equation}
{\eta}\dot{\delta}=\delta A \cos\left(\frac{\Delta\theta_0}{2}\right)\sin(\Delta\theta_0)=2\delta A \cos^2\left(\frac{\Delta\theta_0}{2}\right)\sin\left(\frac{\Delta\theta_0}{2}\right)=2\delta J\cos^2\left(\frac{\Delta\theta_0}{2}\right)
\end{equation}
\begin{equation}
{\eta}\dot{\epsilon}=\epsilon A \cos\left(\frac{\Delta\theta_0}{2}\right)\sin(\Delta\theta_0)=2\epsilon A \cos^2\left(\frac{\Delta\theta_0}{2}\right)\sin\left(\frac{\Delta\theta_0}{2}\right)=2\epsilon J\cos^2\left(\frac{\Delta\theta_0}{2}\right)
\end{equation}
This implies that both the eigenvalues are equal and both are positive since $J>0$ implying that this fixed point is unstable. Thus the only fixed points are at $\Delta\theta=0$ for arbitrary $J$ and $A$ and $\Delta\theta=\pi$ or $-\pi$ when $A>J$.

\subsection{Motile overdamped spins}
Now allow the spins to move in space. For simplicity, we assume that the spins interact with the same strength over arbitrary distances -- a non-metric interaction. The dynamics is described by
\begin{equation}
\dot{s}_1+\frac{\eta}{\chi}s_1=J{\bf v}_1\times{\bf v}_2+A({\bf v}_1\cdot\hat{r}_{12}){\bf v}_1\times{\bf v}_2
\end{equation}
and 
\begin{equation}
\dot{s}_2+\frac{\eta}{\chi}s_2=J{\bf v}_2\times{\bf v}_1-A({\bf v}_2\cdot\hat{r}_{12}){\bf v}_2\times{\bf v}_1.
\end{equation}
The particles also move in the direction they are pointing in:
$
\dot{\bf r}_1={\bf v}_1
$, 
$
\dot{\bf r}_2={\bf v}_2
$
where we have rescaled the particle speed to $1$. This implies that
\begin{equation}
\dot{\hat{r}}_{12}=\frac{{\bf v}_1-{\bf v}_2}{|{\bf v}_1-{\bf v}_2|}=\frac{(\cos\theta_1-\cos\theta_2)\hat{x}+(\sin\theta_1-\sin\theta_2)\hat{y}}{\sqrt{(\cos\theta_1-\cos\theta_2)^2+(\sin\theta_1-\sin\theta_2)^2}}
\end{equation}
when ${\bf v}_1\neq {\bf v}_2$ and $0$ when ${\bf v}_1={\bf v}_2$. Writing ${\bf r}_{12}$ as $R(\cos\Omega,\sin\Omega)$, we obtain the dynamics of $\Omega$ by projecting $\dot{\hat{r}}_{12}$ transverse to $\hat{r}_{12}$ (for $\theta_1\neq\theta_2$)
\begin{equation}
\dot{\Omega}=\frac{-\sin\Omega(\cos\theta_1-\cos\theta_2)+\cos\Omega(\sin\theta_1-\sin\theta_2)}{\sqrt{2-2\cos(\theta_1-\theta_2)}}=\frac{\sin(\theta_1-\Omega)-\sin(\theta_2-\Omega)}{\sqrt{2-2\cos[(\theta_1-\Omega)-(\theta_2-\Omega)]}}
\label{lnjn}
\end{equation}
$\dot{\Omega}=0$ by definition when $\theta_1=\theta_2$. The overdamped equations for the angles are 
\begin{equation}
\label{ang1}
{\eta}\dot{\theta}_1=-J\sin(\theta_1-\theta_2)-A\cos(\theta_1-\Omega)\sin(\theta_1-\theta_2)
\end{equation}
and
\begin{equation}
\label{ang2}
{\eta}\dot{\theta}_2=J\sin(\theta_1-\theta_2)-A\cos(\theta_2-\Omega)\sin(\theta_1-\theta_2)
\end{equation}
Defining the variables $\theta^R_1=\theta_1-\Omega$ and $\theta^R_2=\theta_2-\Omega$, and subtracting $\eta$ times \eqref{lnjn} from \eqref{ang1} and \eqref{ang2}, I get two closed equations for $\theta^R_1$ and $\theta^R_2$:
\begin{equation}
\eta\dot{\theta}^R_1=-J\sin(\theta^R_1-\theta^R_2)-A\cos\theta^R_1\sin(\theta^R_1-\theta^R_2)-\eta\cos\left(\frac{\theta^R_1+\theta^R_2}{2}\right)
\end{equation}
(for $\theta^R_1\neq\theta^R_2$. For $\theta^R_1=\theta^R_2$, the last term vanishes).
\begin{equation}
\eta\dot{\theta}^R_2=J\sin(\theta^R_1-\theta^R_2)-A\cos\theta^R_2\sin(\theta^R_1-\theta^R_2)-\eta\cos\left(\frac{\theta^R_1+\theta^R_2}{2}\right)
\end{equation}
Defining $\Delta\theta=(\theta^R_1-\theta^R_2)/2=(\theta_1-\theta_2)/2$ and $\theta_t=(\theta^R_1+\theta^R_2)/2=(\theta_1+\theta_2-2\Omega)/2$ (note the different definitions from earlier) I finally get
\begin{equation}
\label{sum1}
{\eta}\dot{{\theta}_t}=-2A\sin\Delta\theta\cos\theta_t\cos^2\Delta\theta-\eta\cos\theta_t
\end{equation}
\begin{equation}
\label{diff1}
{\eta}\Delta\dot{{\theta}}=-2\sin\Delta\theta\cos\Delta\theta\left[J-A\sin\theta_t\sin\Delta\theta\right]
\end{equation}
for $\Delta\theta\neq 0$. For $\Delta\theta=0$, the final term in \eqref{sum1} vanishes. Since the final term $\propto \eta$ in the R.H.S. of \eqref{sum1} is not present for $\Delta\theta=0$, it is clear that there is a line of stable fixed points for $\Delta\theta=0$ and arbitrary $\theta_t$. There's also an unstable fixed points with $\Delta\theta=\pi/2$  an $\theta_t=\pi/2$. The linearised equations about this point are
$
\eta\dot{\delta}=\eta\delta
$,
$
\eta\dot{\epsilon}=2\epsilon(J-A)
$.
For $A>J$, the angular difference $\Delta\theta$ is stable, while the $\theta_t$ is unstable. Are there dynamical states with $\Delta\theta=\pi/2$? With $\Delta\theta=\pi/2$ and an arbitrary but fixed $\theta_t=\theta_{t_0}$, $\eta\dot{\theta}_t=-\eta\cos\theta_{t_0}$ and $\eta\Delta\dot{\theta}=0$. The linear perturbation about this with $\theta_t=\theta_{t_0}+\delta$ and $\Delta\theta=\pi/2+\epsilon$ leads to 
\begin{equation}
\eta\dot{\theta}_t=-\eta\cos\theta_{t_0}+\delta\eta\sin\theta_{t_0}
\end{equation}
\begin{equation}
\eta\Delta\dot{\theta}=2\epsilon(J-A\sin\theta_{t_0})
\end{equation}
\begin{figure}
 \includegraphics[width=7cm]{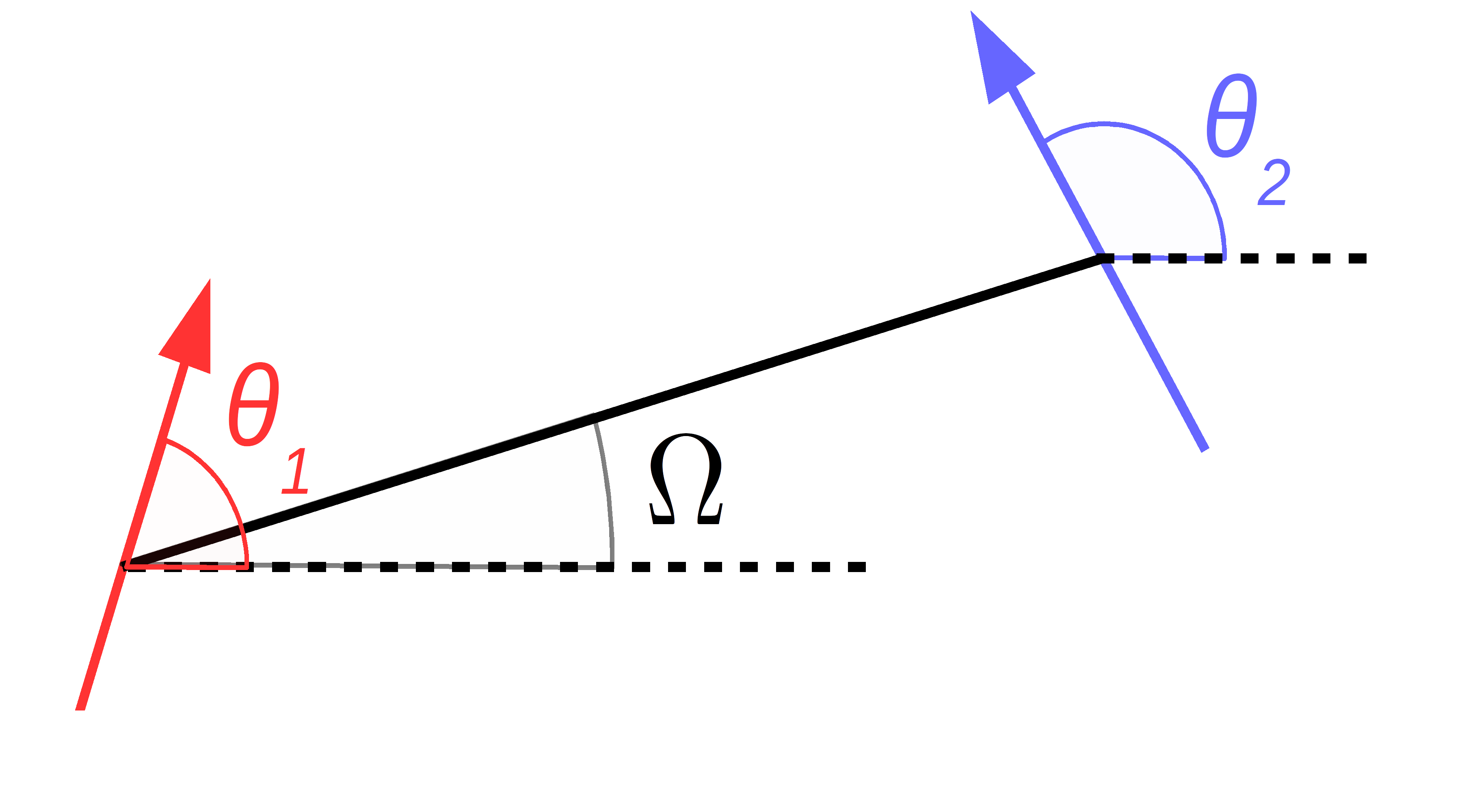}
\caption{Two motile spins making angles $\theta_1$ and $\theta_2$ measured from a fiducial line with the angle of the line joining the two spins $\Omega$ also measured from the same fiducial line.} 
\label{twopart}
\end{figure}
Thus, when $\sin\theta_{t_0}<0$ {\it and} $A\sin\theta_{t_0}>J$, perturbations decay to the state with steadily increasing (at a constant rate) $\theta_t$ with $\Delta\theta=\pi/2$, i.e., with spins pointing opposite to each other. However, recall that $\theta_t=(\theta_1+\theta_2-2\Omega)/2$. Thus, this state does not correspond to a spontaneous rotation of the two spins, but two spins pointing away from each other and flying away from each other. When two spins fly away from each other at constant speeds, if they are not collinear, the angle that a line joining them makes with a fiducial line, increases continuously. This is the situation that this corresponds to. This is trivial since two spins flying away from each other would stop interacting if a distance cut-off were implemented.

Further, for $|A|>J$, there is a fixed point for $\theta_t=\pi/2$ and $\sin\Delta\theta=J/A$. This fixed point is still always unstable since $J>0$ and $\eta>0$. Finally, there is another fixed point for
\begin{equation}
\sin \Delta\theta_0\cos^2\Delta\theta_0=-\frac{\eta}{2A};\,\,\,\,\,\,\,\,  \sin\theta_{t_0}\sin\Delta\theta_0=\frac{J}{A}.
\end{equation}
The first condition requires $A>\eta$ and the second $A>J$. Furthermore, it is clear from the first condition that $\sin\Delta\theta_0$ has a sign opposite to $A$ and from the second condition, $\sin\theta_{t_0}$ is always negative.  For the final fixed point, the linearised equations are
\begin{equation}
\eta\dot{\delta}=\epsilon A\cos(\Delta\theta_0)(1-3\cos(2\Delta\theta_0)\cos\theta_{t_0}
\label{deleqn}
\end{equation}
\begin{equation}
\eta\dot{\epsilon}=A\sin(2\Delta\theta_0)[\sin\Delta\theta_0\cos\theta_{t_0}\delta+\cos\Delta\theta_0\sin\theta_{t_0}\epsilon]
\label{epseqn}
\end{equation}
or 
\begin{equation}
\eta\begin{pmatrix}\dot{\delta}\\\dot{\epsilon}\end{pmatrix}=\begin{pmatrix}0 & A\cos(\Delta\theta_0)(1-3\cos(2\Delta\theta_0)\cos\theta_{t_0}\\ A\sin(2\Delta\theta_0)\sin\Delta\theta_0\cos\theta_{t_0} &2A\sin\Delta\theta_0\cos^2\Delta\theta_0\sin\theta_{t_0}\end{pmatrix}\begin{pmatrix}\delta\\\epsilon\end{pmatrix}
\end{equation}
Since $A\sin\Delta\theta_0\sin\theta_{t_0}=J>0$, $2A\sin\Delta\theta_0\cos^2\Delta\theta_0\sin\theta_{t_0}>0$. Since the trace of the matrix is the sum of the two eigenvalues, and since this implies that the trace is positive, it immediately implies that at least one of the two eigenvalues is positive. Thus, this fixed point is also always unstable. Thus, both for fixed and motile spins we find that when $J>A$, two spins always align. When $J<A$, spins which are behind each other tend to anti-align and therefore, move away from each other, which when a distance cut-off is implemented implies that they quickly stop interacting.

\section{Dynamical theory of motile polar rods, with spin angular momenta, on a substrate}
\label{Pol_rods}
In this appendix we construct the dynamical equations of motion for a system of motile polar rods on a substrate, explicitly retaining their spin angular momenta. The dynamical fields are the density $\rho({\bf x},t)$, the polarisation ${\bf v}({\bf x}, t)$, the velocity ${\bf u}({\bf x}, t)$, and the spin angular momentum ${\bf s}({\bf x}, t)$ \cite{Stark}. 
The conserved equation for the density field is 
\begin{equation}
\label{denseqn}
D_t\rho=-\rho\nabla\cdot{\bf u}.
\end{equation}
The polarisation field has the dynamical equation
\begin{equation}
\label{poleqn}
D_t{\bf v}=\frac{1}{\chi}{\bf s}\times {\bf v}-\Gamma_v{\bf h}+\frac{\lambda_1}{\chi}\nabla\times{\bf s}-\frac{\Gamma'_u}{|v|^2}{\bf v}({\bf v}\cdot{\bf u}),
\end{equation}
where the first term denotes the rotation of the polarisation with the angular frequency ${\bf s}/\chi$, the second relaxation to the minimum of a standard Landau-de Gennes free energy for polar liquid crystals \cite{kung} 
\begin{equation}
F=\int d^d{\bf r}\left[-\frac{\alpha}{2}{\bf v}\cdot{\bf v}+\frac{\beta}{4}({\bf v}\cdot{\bf v})^2+\frac{K}{2}(\nabla{\bf v})^2+K_p({\bf v}\cdot{\bf v})\nabla\cdot{\bf v}+\gamma\rho\nabla\cdot{\bf v}\right]
\end{equation}
with ${\bf h}=\delta F/\delta{\bf p}$, the third term is a second passive reactive coupling between ${\bf s}$ and ${\bf v}$, and the final term implies that the magnitude of polarisation changes with imposed flow. Since this implies implies that the polarisation field can respond to a mean velocity and not only its gradient, it is only allowed in systems that do not conserve momentum. The equation of motion for the spin is
\begin{equation}
\label{spneqn}
D_t{\bf s}=-{\bf v}\times{\bf h}-\lambda_1\nabla\times{\bf h}-\Gamma_u({\bf v}\times{\bf u})-\Gamma_\omega\left(\frac{{\bf s}}{\chi}-\boldsymbol{\omega}\right)+\Gamma_D{\bf v}\times({\bf v}\cdot\bsf{D})+\Gamma_D'{\bf v}\times(\nabla\cdot\bsf{D})-\frac{\eta-\Gamma_\omega}{\chi}{\bf s},
\end{equation}
where the first two terms are the reactive Onsager partners to the terms involving ${\bf s}/\chi$ in \eqref{poleqn}, $\Gamma_u$, $\Gamma_\omega$, $\Gamma_D$ and $\Gamma'_D$ are dissipative cross-couplings with the velocity field ${\bf u}$, where $D_{ij}=(\partial_iu_j+\partial_ju_i)/2$ and $\boldsymbol{\omega}=(\nabla\times{\bf u})/2$, and the final dissipative term is due to the friction with the substrate. The terms proportional to $\Gamma_u$ and $\eta$ are only allowed in systems that do not conserve momentum and angular momentum.
\begin{equation}
\label{flveleqn}
\rho D_t {\bf u}=-\Gamma{\bf u}+\zeta{\bf v}+\frac{\Gamma_u}{\chi}{\bf v}\times{\bf s}+\frac{\Gamma'_u}{|v|^2}{\bf vv}\cdot{\bf h}-\rho\nabla\frac{\delta F}{\delta\rho}+\nabla\cdot\boldsymbol{\sigma}
\end{equation}
where $\boldsymbol{\sigma}$ contains standard Onsager symmetric partners of the terms $\Gamma_\omega$, $\Gamma_D$ and $\Gamma'_D$ in \eqref{spneqn} \cite{Stark}. Since these terms are all subdominant to those retained in \eqref{flveleqn}, and will be ignored in what follows, we do not explicitly write them here. 

To check the consistency of these equations, let us first eliminate ${\bf s}$ in terms of the other fields to obtain
\begin{equation}
{\bf s}\approx-\frac{\chi}{\eta}\left[{\bf v}\times{\bf h}+\lambda_1\nabla\times{\bf h}+\Gamma_u({\bf v}\times{\bf u})\right]
\end{equation}
ignoring all gradient terms in the fluid velocity.
This, when replaced in the velocity and the polarisation equations, yields
\begin{equation}
\label{polrelsp}
D_t{\bf v}=\frac{1}{\eta}{\bf v}\times {\bf v}\times{\bf h}-\Gamma_v{\bf h}+\frac{\Gamma_u}{\eta}{\bf v}\times ({\bf v}\times{\bf u})-\frac{\Gamma'_u}{|v|^2}{\bf v}({\bf v}\cdot{\bf u})+\frac{\lambda_1\Gamma_u}{\eta}\nabla\times({\bf v}\times{\bf u})
\end{equation}
\begin{equation}
\label{velrelsp}
\Gamma{\bf u}+\frac{\Gamma_u^2}{\eta}{\bf v}\times({\bf v}\times{\bf u})=\zeta{\bf v}-\frac{\Gamma_u}{\eta}{\bf v}\times({\bf v}\times{\bf h})+\frac{\Gamma'_u}{|v|^2}{\bf vv}\cdot{\bf h}+\frac{\Gamma_u\lambda_1}{\eta}{\bf v}\times(\nabla\times{\bf h})-\rho\nabla\frac{\delta F}{\delta\rho}+...
\end{equation}
These are the equations of motion for a polar motile fluid on a substrate \cite{Lauga, Ano_pol, Harsh, LPDJSTAT}. Specifically, the two terms $\Gamma_v {\bf h}$ and $(1/\eta){\bf v}\times{\bf v}\times{\bf h}$ in \eqref{polrelsp} together constitute an anisotropic relaxation of the polarisation to the minimum of the potential, the two terms $({\Gamma_u}/{\eta}){\bf v}\times ({\bf v}\times{\bf u})-({\Gamma'_u}/{|v|^2}){\bf v}({\bf v}\cdot{\bf u})$ and their Onsager anti-symmetric counterparts in \eqref{velrelsp} together constitute an anisotropic analogue of the ``weathercock'' term in \cite{Ano_pol} (which has an isotropic coefficient $\Lambda$ there), and the term $\propto \nabla\times({\bf v}\times{\bf u})$ is a higher order Onsager anti-symmetric coupling between the polarisation and the velocity (which would also be forbidden in a momentum-conserved system). The terms on the L.H.S. of \eqref{velrelsp} constitute an anisotropic friction and $\zeta{\bf v}$ is the motility (which is the only source of activity in this model) as discussed earlier.

We now look at the opposite situation -- when the velocity field is eliminated but the spin angular momentum is retained -- as appropriate for inertial spin models. To lowest order in gradients,
\begin{equation}
{\bf u}\approx\frac{\zeta}{\Gamma}{\bf v}+\frac{\Gamma_u}{\chi\Gamma}{\bf v}\times{\bf s}+\frac{\Gamma'_u}{\Gamma|v|^2}{\bf vv}\cdot{\bf h}
\end{equation}
First, using this in \eqref{denseqn} we get
\begin{equation}
\partial_t\rho+\frac{\zeta}{\Gamma}{\bf v}\cdot\nabla\rho=-\rho\frac{\zeta}{\Gamma}\nabla\cdot{\bf v}-\frac{\Gamma_u}{\chi\Gamma}\nabla\cdot({\bf v}\times{\bf s})
\end{equation}
where we, anticipating slightly, have only retained terms that will have any contribution at the linear order and have not retained a term with ${\bf h}$ since in the ordered phase, ${\bf h}$ is subdominant to ${\bf v}$.
F/rom \eqref{poleqn}, we obtain
\begin{equation}
\label{poleqn}
\partial_t{\bf v}+\frac{\zeta}{\Gamma}{\bf v}\cdot\nabla{\bf v}=\frac{1}{\chi}{\bf s}\times {\bf v}-\Gamma_v{\bf h}-\frac{\Gamma'^2_u}{\Gamma|v|^2}{\bf v}({\bf v}\cdot{\bf h})+\frac{\lambda_1}{\chi}\nabla\times{\bf s}-\frac{\Gamma'_u\zeta}{\Gamma}{\bf v},
\end{equation}
where we have again only retained terms that have any contribution at the linear order, and from which we see that at this stage, both the density and the polarisation fields are advected with the same speed. Finally, the equation of motion for the spin variable is  
\begin{multline}
\label{spingen}
\partial_t{\bf s}+\frac{\zeta}{\Gamma}{\bf v}\cdot\nabla{\bf s}=-{\bf v}\times{\bf h}-\frac{\Gamma^2_u}{\chi\Gamma}{\bf v}\times({\bf v}\times{\bf s})+\frac{\Gamma_\omega\zeta}{2\Gamma}\nabla\times{\bf v}+\frac{\Gamma_\omega\Gamma_u}{2\chi\Gamma}\nabla\times({\bf v}\times{\bf s})+\frac{\Gamma_D\zeta}{2\Gamma}{\bf v}\times\left({\bf v}\cdot\nabla{\bf v}+\frac{1}{2}\nabla|v^2|\right)\\+\frac{\Gamma'_D\zeta}{2\Gamma}{\bf v}\times[\nabla^2{\bf v}+\nabla(\nabla\cdot{\bf v})]+\frac{\Gamma_D\Gamma_u}{2\chi\Gamma}{\bf v}\times[{\bf v}\cdot\nabla({\bf v}\times{\bf s})]-\frac{\eta}{\chi}{\bf s}
\end{multline}
where we have only retained terms that will contribute to the linear equations, ignored terms that are higher order in gradients of ${\bf h}$ and terms that are second order in gradients of ${\bf s}$. This is a generalised version of the model that we obtained earlier from the microscopic theory. Specifically, the term that was characteristic of antisymmetric \SR{coupling} there, $\propto {\bf v}\times[{\bf v}\cdot\nabla{\bf v}]$ in the $\dot{{\bf s}}$ equation, arises here from terms that in passive system would lead to flow alignment. The symmetric \SR{terms} arise both from the free energy and a higher order, polar flow alignment term. All the terms on the R.H.S. of \eqref{spingen} containing ${\bf s}$ are dissipative, with the ones $\propto \Gamma^2_u/\chi\Gamma$ and $\eta/\chi$ leading to an anisotropic friction and the remaining ones to dissipative terms at first order in gradients in ${\bf s}$. This may seem somewhat counter-intuitive; after all, the non-frictional dissipative terms are expected to be $\sim\nabla^2$. However, in polar systems on substrates, the momentum field and the angular momentum fields can both have dissipative terms $\sim {\bf v}\cdot\nabla\delta F/{\delta {\bf g}}$ and $\sim {\bf v}\cdot\nabla\delta F/{\delta {\bf s}}$ where $F$ is the free energy including the kinetic terms which in the ordered polar phase leads to linear terms $\sim\partial_\parallel{\bf u}$ in the velocity equation and $\sim \partial_\parallel{\bf s}$ in the spin angular momentum equation. The first order in gradient dissipative terms in ${\bf s}$ in \eqref{spingen} are essentially of this form.

To underscore the equivalence of this formulation with the one starting from a model with microscopic antisymmetric \SR{aligning torques}, we now linearise the dynamical equations about a state with ${\bf s}=0$, ${\bf v}=v_0\hat{x}$, with $v_0=\sqrt{(\alpha+\Gamma'_u\zeta/\Gamma)/\beta}$, ignoring all anisotropies in the spin friction and the dissipative kinetic coefficient for the polarisation field, ignoring the first order in gradient dissipative terms in ${\bf s}$,  transforming to a frame moving with the mean flock velocity and defining $\bar{\rho}_0\equiv\rho_0\zeta/\Gamma$
\begin{equation}
\label{lindens}
\partial_t\delta\rho=-\bar{\rho}_0\partial_\perp\delta v_\perp+\frac{\Gamma_uv_0}{\chi\Gamma}\partial_\perp s
\end{equation}
\begin{equation}
\label{linpol1}
\partial_t\delta v_\perp=\frac{v_0}{\chi}s+\Gamma_vK\nabla^2\delta v_\perp-\Gamma_v\gamma\partial_\perp\delta\rho-\frac{\lambda}{\chi}\partial_\parallel s
\end{equation}
\begin{equation}
\label{linspn1}
\partial_ts=-\frac{\eta}{\chi}s+\left(K+\frac{\Gamma'_D\zeta v_0}{2\Gamma}\right)\nabla^2\delta v_\perp+\frac{(\Gamma_\omega+\Gamma_D v_0^2)\zeta}{2\Gamma}\partial_\parallel\delta v_\perp+\frac{\Gamma'_D\zeta}{2\Gamma}\partial_\perp^2\delta v_\perp
\end{equation}
This model is a more general variant of the model with intrinsic antisymmetric \SR{coupling} and indeed, the antisymmetric \SR{term} induces a longitudinal instability, when its coefficient ${(\Gamma_\omega+\Gamma_D v_0^2)\zeta}/{2\Gamma}$ is large enough. Indeed, the frequencies for the two modes along $k_\parallel$ (the density equation is decoupled for perturbations in direction) are, to leading order in wavevectors,
\begin{equation}
\omega_1=-i\frac{\eta}{\chi}+\frac{v_0}{\eta}\frac{(\Gamma_\omega+\Gamma_D v_0^2)\zeta}{2\Gamma}k_\parallel
\end{equation}
\begin{equation}
\omega_2=-\frac{v_0}{\eta}\frac{(\Gamma_\omega+\Gamma_D v_0^2)\zeta}{2\Gamma}k_\parallel-\left[\Gamma_vK+\left(K+\frac{\Gamma'_D\zeta v_0}{2\Gamma}\right)\frac{v_0}{\eta}-\frac{1}{\eta^3}\frac{(\Gamma_\omega+\Gamma_D v_0^2)\zeta}{2\Gamma}\left(\lambda\eta^2+\chi v_0^2\frac{(\Gamma_\omega+\Gamma_D v_0^2)\zeta}{2\Gamma}\right)\right]k_\parallel^2
\end{equation}
The term in the square bracket for $\omega_2$ turns negative due to the flow alignment terms in this formulation. Note that in the model with microscopic anisotropy, we ignored a term $\propto \nabla\times{\bf s}$ in the $\dot{{\bf v}}$ equation. 

\section{\AMN{Elimination of the fast spin variable}}
\label{ElimSpin}
\AMN{To eliminate the fast spin variable, which only couples to the velocity, we start with the coupled equations
\begin{equation}
\label{appveleq}
D_t{\bf v}=-\frac{1}{\chi}{\bf v}\times {\bf s}-\Gamma_v\frac{\delta U}{\delta {\bf v}}-\frac{1}{\rho}\nabla f(\rho)
\end{equation}
and
\begin{equation}
\frac{\eta}{\chi}\bigg(1+\frac{\chi}{\eta}D_t\bigg){\bf s}=\frac{J}{v_0^2}{\bf v}\times\nabla^2{\bf v}+\frac{\mathcal{A}}{v_0^3}{\bf v}\times({\bf v}\cdot\nabla){\bf v}+\frac{J_A}{v_0^4}{\bf v}\times({\bf v}\cdot\nabla)^2{\bf v}
\end{equation}
Then, a formal solution of ${\bf s}$ has the form
\begin{equation}
{\bf s}=\frac{\chi}{\eta}\left[1+\sum^\infty_{n=1}\left(\frac{-\chi}{\eta}\right)^nD^n_t\right]\bigg[\frac{\mathcal{A}}{v_0^3}{\bf v}\times({\bf v}\cdot\nabla{\bf v})+\frac{J}{v_0^2}{\bf v}\times\nabla^2{\bf v}+\frac{J_A}{v_0^4}{\bf v}\times[({\bf v}\cdot\nabla)^2{\bf v}]\bigg]
\end{equation}
which, when inserted into \eqref{appveleq} yields
\begin{equation}
D_t{\bf v}=-\Gamma_v\frac{\delta U}{\delta {\bf v}}-\frac{1}{\rho}\nabla f(\rho)-\frac{1}{\eta}{\bf v}\times\left[\left\{1+\sum^\infty_{n=1}\left(\frac{-\chi}{\eta}\right)^nD^n_t\right\}\bigg(\frac{J}{v_0^2}{\bf v}\times\nabla^2{\bf v}+\frac{\mathcal{A}}{v_0^3}{\bf v}\times({\bf v}\cdot\nabla){\bf v}+\frac{J_A}{v_0^4}{\bf v}\times({\bf v}\cdot\nabla)^2{\bf v}\bigg)\right]
\end{equation}
We now solve for $D_t{\bf v}$ perturbatively in $1/\eta$ and only nonlinear retain terms with at most two fields and one gradient and linear terms with at most two gradients. To zeroth order in $1/\eta$, 
\begin{equation}
D_t{\bf v}=-\Gamma_v\frac{\delta U}{\delta {\bf v}}-\frac{1}{\rho}\nabla f(\rho)
\end{equation}
and defining ${\bsf T}={\bsf I}-\hat{{\bf v}}\hat{{\bf v}}$, $D_t{\bf v}$ to first order in $1/\eta$ reads
\begin{equation}
D_t{\bf v}=-\Gamma_v\frac{\delta U}{\delta {\bf v}}-\frac{1}{\rho}\nabla f(\rho)+\frac{1}{\eta}{\bsf T}\cdot\bigg({J}\nabla^2{\bf v}+\frac{\mathcal{A}}{v_0}({\bf v}\cdot\nabla){\bf v}+\frac{J_A}{v_0^2}({\bf v}\cdot\nabla)^2{\bf v}\bigg)
\end{equation}
To write the $\mathcal{O}(1/\eta^2)$ contribution, we note that to zeroth order in gradient, $-\delta U/\delta{\bf v}\approx -2\delta v_\parallel\hat{{\bf v}}_0$ where $\hat{{\bf v}}_0$ is the unit vector along the ordering direction. Then, expanding the term
\begin{equation}
\frac{\chi}{\eta^2}{\bf v}\times D_t \bigg(\frac{J}{v_0^2}{\bf v}\times\nabla^2{\bf v}+\frac{\mathcal{A}}{v_0^3}{\bf v}\times({\bf v}\cdot\nabla){\bf v}+\frac{J_A}{v_0^4}{\bf v}\times({\bf v}\cdot\nabla)^2{\bf v}\bigg)
\end{equation}
and replacing every instance of $D_t{\bf v}$ by its zeroth order in $1/\eta$ value, we obtain the $\mathcal{O}(1/\eta^2)$ correction, restricting ourselves to linear terms with two gradients and nonlinear ones with a single gradient and two fields
\begin{equation}
\frac{2\chi\Gamma_v}{\eta^2}{\bsf T}\cdot \bigg({J}{}\nabla^2\delta v_\parallel\hat{{\bf v}}_0+\frac{\mathcal{A}}{v_0}({\bf v}\cdot\nabla)\delta v_\parallel\hat{{\bf v}}_0+\frac{\mathcal{A}}{v_0}(\delta v_\parallel\partial_\parallel){\bf v}+\frac{J_A}{v_0^2}({\bf v}\cdot\nabla)^2\delta v_\parallel\hat{{\bf v}}_0\bigg)+\frac{2\chi\mathcal{A}f'(\rho_0)}{\eta^2\rho_0}{\bsf T}\cdot(\partial_\parallel\nabla)\delta\rho
\end{equation}
where the prime denotes differentiation with $\rho$ and $\rho_0$ is the steady-state density with $\rho=\rho_0+\delta\rho$. To find the $\mathcal{O}(1/\eta^3)$ correction, we need to replace every instance of $D^2_t{\bf v}$ in the term
\begin{equation}
-\frac{\chi^2}{\eta^3}{\bf v}\times D^2_t\bigg(\frac{J}{v_0^2}{\bf v}\times\nabla^2{\bf v}+\frac{\mathcal{A}}{v_0^3}{\bf v}\times({\bf v}\cdot\nabla){\bf v}+\frac{J_A}{v_0^4}{\bf v}\times({\bf v}\cdot\nabla)^2{\bf v}\bigg)
\end{equation}
with its zeroth order in $1/\eta$ value and $D_t{\bf v}$ in the term
\begin{equation}
\frac{\chi}{\eta^2}{\bf v}\times D_t \bigg(\frac{J}{v_0^2}{\bf v}\times\nabla^2{\bf v}+\frac{\mathcal{A}}{v_0^3}{\bf v}\times({\bf v}\cdot\nabla){\bf v}+\frac{J_A}{v_0^4}{\bf v}\times({\bf v}\cdot\nabla)^2{\bf v}\bigg)
\end{equation}
with the $\mathcal{O}(1/\eta)$ correction to $D_t{\bf v}$. To zeroth order in $1/\eta$ and to linear order,
\begin{equation}
D_t^2{\bf v}=-\Gamma_uD_t\frac{\delta U}{\delta{\bf v}}-\frac{f'(\rho_0)}{\rho_0}\nabla D_t\delta\rho\approx-2\Gamma_uD_t\delta v_\parallel\hat{{\bf v}}_0+{f'(\rho_0)}\nabla\nabla\cdot{\bf v}\approx \left(4\Gamma_u^2\delta v_\parallel+\frac{f'(\rho_0)}{\rho_0}\partial_\parallel\delta\rho\right)\hat{{\bf v}}_0
\end{equation}
where we have only retained the lowest order in gradient terms since these are the only ones that will contribute to the final expression for $D_t{\bf v}$ to $\mathcal{O}(1/\eta^3)$ up to the order in gradients and fields we re interested in. When this expression for $D_t^2{\bf v}$ is used, we obtain the term
\begin{equation}
\frac{\chi^2}{\eta^3}{\bsf T}\cdot\left[\bigg({4\Gamma_u^2J}\nabla^2\delta v_\parallel+\frac{4\Gamma_u^2\mathcal{A}}{v_0}({\bf v}\cdot\nabla)\delta v_\parallel+\frac{4\Gamma_u^2J_A}{v_0^2}({\bf v}\cdot\nabla)^2\delta v_\parallel+\frac{f'(\rho_0)\mathcal{A}}{\rho_0}\partial_\parallel^2\delta\rho\bigg)\hat{{\bf v}}_0+\frac{4\Gamma_u^2\mathcal{A}}{v_0}(\delta v_\parallel\partial_\parallel){\bf v}\right].
\end{equation}
Finally, by replacing $D_t{\bf v}$ with its $\mathcal{O}(1/\eta)$ expression, we obtain only one term to second order in gradients:
\begin{equation}
-\frac{\chi\mathcal{A}^2}{\eta^3 v_0^2}{\bsf T}\cdot({\bf v}\cdot\nabla)^2{\bf v}
\end{equation}
Putting all of these together, the expression for $D_t{\bf v}$ is 
\begin{multline}
D_t{\bf v}=-\Gamma_v\frac{\delta U}{\delta {\bf v}}-\frac{1}{\rho}\nabla f(\rho)+\frac{1}{\eta}{\bsf T}\cdot\bigg({J}\nabla^2{\bf v}+\frac{\mathcal{A}}{v_0}({\bf v}\cdot\nabla){\bf v}+\frac{J_A}{v_0^2}({\bf v}\cdot\nabla)^2{\bf v}\bigg)\\\frac{2\chi\Gamma_v}{\eta^2}{\bsf T}\cdot \bigg({J}{}\nabla^2\delta v_\parallel\hat{{\bf v}}_0+\frac{\mathcal{A}}{v_0}({\bf v}\cdot\nabla)\delta v_\parallel\hat{{\bf v}}_0+\frac{\mathcal{A}}{v_0}(\delta v_\parallel\partial_\parallel){\bf v}+\frac{J_A}{v_0^2}({\bf v}\cdot\nabla)^2\delta v_\parallel\hat{{\bf v}}_0\bigg)+\frac{2\chi\mathcal{A}f'(\rho_0)}{\eta^2\rho_0}{\bsf T}\cdot(\partial_\parallel\nabla)\delta\rho\\\frac{\chi^2}{\eta^3}{\bsf T}\cdot\left[\bigg({4\Gamma_u^2J}\nabla^2\delta v_\parallel+\frac{4\Gamma_u^2\mathcal{A}}{v_0}({\bf v}\cdot\nabla)\delta v_\parallel+\frac{4\Gamma_u^2J_A}{v_0^2}({\bf v}\cdot\nabla)^2\delta v_\parallel+\frac{f'(\rho_0)\mathcal{A}}{\rho_0}\partial_\parallel^2\delta\rho\bigg)\hat{{\bf v}}_0+\frac{4\Gamma_u^2\mathcal{A}}{v_0}(\delta v_\parallel\partial_\parallel){\bf v}\right]-\frac{\chi\mathcal{A}^2}{\eta^3 v_0^2}{\bsf T}\cdot({\bf v}\cdot\nabla)^2{\bf v}
\end{multline}
This is the analogue of the Toner-Tu equation that is obtained when the spin angular momentum is eliminated. 
Note that $\delta v_\parallel$ is not a slow variable. In fact, for the $U$ we have chosen, $\delta v_\parallel\sim\delta v_\perp^2$ (i.e., it doesn't depend on density fluctuations). Therefore, all the nonlinear terms involving $\delta v_\parallel$ are subdominant to $\delta v_\perp\partial_\perp\delta v_\perp$. Therefore, projecting this equation transverse to the mean velocity, we obtain \eqref{TonTulike} of the main text where we retain linear term till second order in gradients and nonlinear terms with two slow fields and one gradient. Note that while the equation for $\partial_t v_\perp$ was obtained through an expansion in $1/\eta$, at this order in gradients and fields, it is correct to all orders in $1/\eta$. This is because it can be easily checked that no higher order $1/\eta$ contribution can lead to a term at this order in gradients and fields in the equation for $D_tv_\perp$.}

\section{Appendix: Mode structure for the conserved density model}
\label{cons_dens}
As discussed in the main text, the equations of motion for a model with \SR{asymmetric} \SR{coupling} and a conserved density is 
        \begin{equation}
        \label{vperpeq1}
        \partial_t \delta v_\bot = -\sigma\partial_\bot \delta\rho + 
\frac{v_0}{\chi}s+\bar{K}\nabla^2\delta v_\perp
        \end{equation}
where $\bar{K}=\Gamma_v K$,
        \begin{equation}
\label{spinlineq1}
        \partial_t s = {J\over v_0}\nabla^2\delta v_\perp+{J_A\over v_0}\partial_\parallel^2\delta v_\perp + {\mathcal{A} 
\over v_0} \partial_\parallel\delta v_\perp - \frac{\eta}{\chi}s
        \end{equation}
      \begin{equation}
      \partial_t\delta\rho = -\rho_0\partial_\perp\delta v_\perp.
      \end{equation}
This implies the mode structure
\begin{equation}
\omega_1=-i\frac{\eta}{\chi}+\frac{\mathcal{A}}{\eta}k_\parallel
\end{equation}
\begin{multline}
\omega_2=-\frac{\mathcal{A}}{2\eta}k_\parallel+\frac{\sqrt{\mathcal{A}^2k_\parallel^2+4\eta^2\rho_0\sigma k_\perp^2}}{2\eta}\\+\frac{i}{2\eta^3}\left[-(J+\bar{K}\eta)\eta^2k^2-(J_A\eta^2-\mathcal{A}^2\chi) k_\parallel^2+\mathcal{A}k_\parallel\frac{(J+\bar{K}\eta)\eta^2k^2+(J_A\eta^2-\mathcal{A}^2\chi) k_\parallel^2-2\rho_0\eta^2\chi\sigma k_\perp^2}{\sqrt{\mathcal{A}^2k_\parallel^2+4\eta^2\rho_0\sigma k_\perp^2}}\right]
\end{multline}
\begin{multline}
\omega_3=-\frac{\mathcal{A}}{2\eta}k_\parallel-\frac{i\sqrt{\mathcal{A}^2k_\parallel^2+4\eta^2\rho_0\sigma k_\perp^2}}{2\eta}\\+\frac{i}{2\eta^3}\left[-(J+\bar{K}\eta)\eta^2k^2-(J_A\eta^2-\mathcal{A}^2\chi) k_\parallel^2-\mathcal{A}k_\parallel\frac{(J+\bar{K}\eta)\eta^2k^2+(J_A\eta^2-\mathcal{A}^2\chi) k_\parallel^2-2\rho_0\eta^2\chi\sigma k_\perp^2}{\sqrt{\mathcal{A}^2k_\parallel^2+4\eta^2\rho_0\sigma k_\perp^2}}\right]
\end{multline}
From this, we immediately see that for wavevectors along the ordering direction, $\omega_2$ vanishes and $\omega_1$ and $\omega_3$ yield the mode structure derived for a model without density fluctuations in the main text for $k_\parallel$. This is because along $k_\parallel$, the denisty equation decouples and, since we had not included a diffusive term in the density equation, its relaxation rate vanishes.


\begin{thebibliography}{99}
\bibitem{air} The dynamics of the air, which does not appear explicitly in our 
treatment, can be taken to balance the books in this regard, as with linear 
momentum in standard ``dry'' flocking models \cite{RMP}.
\bibitem{behav} It should be kept in mind also that the torque is 
behavioural, not mechanical, and that the birds could conceivably switch 
properties such as the antisymmetry on or off according to circumstances.
\bibitem{comm_diff} \AMR{If we instead consider a collection of non-motile but spatially diffusing polar particles with asymmetric spin interactions, general symmetry arguments suggest that density fluctuations will be advected by the polarisation as in the Toner-Tu theory of motile polar particles.}
\end{thebibliography}
\end{document}